\documentclass[twocolumn,  twocolappendix]{aastex631}
\usepackage{graphicx,amsfonts,color,comment,amsmath,hyperref,float}
\usepackage{dcolumn}
\usepackage{bm}
\usepackage{url}
\usepackage{soul}

\begin{document}

\title{The TeV Diffuse Cosmic Neutrino Spectrum and the Nature of Astrophysical Neutrino Sources \\} 
 
\author{Ke Fang}
\affiliation{Department of Physics, Wisconsin IceCube Particle Astrophysics Center, University of Wisconsin, Madison, WI, 53706}

\author{John S. Gallagher}
\affiliation{
Department of Astronomy,  University of Wisconsin, Madison, WI, 53706
}

\author{Francis Halzen}
\affiliation{Department of Physics, Wisconsin IceCube Particle Astrophysics Center, University of Wisconsin, Madison, WI, 53706}

\date{\today}

\begin{abstract}
The diffuse flux of cosmic neutrinos has been measured by the IceCube Observatory from TeV to PeV energies. We show that an improved characterization of this flux at the lower energies, TeV and sub-TeV, reveals important information on the nature of the astrophysical neutrino sources in a model-independent way. Most significantly, it could confirm the present indications that neutrinos originate in cosmic environments that are optically thick to GeV-TeV $\gamma$-rays. This conclusion will become inevitable if an uninterrupted or even steeper neutrino power law is observed in the TeV region. In such $\gamma$-ray-obscured sources, the $\gamma$-rays that inevitably accompany cosmic neutrinos will cascade down to MeV-GeV energies. The requirement that the cascaded $\gamma$-ray flux accompanying cosmic neutrinos should not exceed the observed diffuse $\gamma$-ray background, puts constraints on the peak energy and density of the radiation fields in the sources. Our calculations inspired by the existing data suggest that a fraction of the observed diffuse MeV-GeV $\gamma$-ray background may be contributed by neutrino sources with intense radiation fields that obscure the high-energy $\gamma$-ray emission accompanying the neutrinos.   
\end{abstract}


\section{Introduction}

A high-energy all-sky neutrino flux of astrophysical origin predominantly of extragalactic origin has been detected by the IceCube Observatory \citep{PhysRevLett.113.101101}. Several independent measurements of the diffuse cosmic neutrino spectrum have been made using neutrino event samples obtained by a variety of selection criteria. The spectrum is consistent with a single power-law, $dN/dE_\nu \propto E_\nu^{-\gamma_{\rm astro}}$, with an index $\gamma_{\rm astro}\sim 2.4-2.9$ as summarized in Table~\ref{table:parameters} \citep{inelasticity5yr, HESE75yr, cascade6yr, numu95yr}. Other spectral models, including a double power-law model with a hard and soft component, have been fit to the IceCube data samples. No convincing indication of an additional component has been found. A mild excess above the atmospheric backgrounds with a similar index is also observed in the ANTARES data  \citep{Fusco:2019mN}.

Using 10 years of IceCube data obtained with the completed detector, the time-integrated search for individual sources contributing to the diffuse flux reveals evidence for an anisotropy on the sky contributed by four potential astrophysical neutrino sources \citep{sourceSearch10yr}. Three out of the four have a spectral index $\gamma_{\rm astro}\gtrsim 3.0$. In addition, the energy flux of neutrinos from the most significant source, NGC~1068, is found to be much higher than that of $\gamma$ rays, indicating that the $\gamma$ rays accompanying the neutrinos are attenuated in the environment where they are produced.

High-energy $\gamma$-ray opacity predominantly arises from interactions with background photons via two-photon annihilation: pair production ($\gamma\gamma_b\rightarrow e^+e^-$) and double pair production ($\gamma\gamma_b\rightarrow e^+e^-e^+e^-$). The final-state electrons and positrons up-scatter background photons via inverse Compton scatter ($e\gamma_b\rightarrow e\gamma$) and triplet pair production ($e\gamma_b\rightarrow ee^+e^-$). The $\gamma$-ray and pair products initiate electromagnetic cascades in which they continue to interact through the same processes until their energy falls below the interaction threshold. Depending on the optical depth of the source environment, the cascade may be initiated either inside the source or during the propagation in the extragalactic background light (EBL) to our detectors. In the former case, the source radiation field can be optically thick to $\gamma$-rays of GeV to PeV energies and optically thick cases will be referred to as ``$\gamma$-ray-obscured" below. The cascade $\gamma$-rays from neutrino sources may show up at MeV to GeV energies, depending on the radiation background of the neutrino sources, and contribute to the diffuse $\gamma$-ray background. In the latter case,  $\gamma$-rays may escape from the sources without significant attenuation and produce electromagnetic cascades in the EBL with energies ranging from GeV to TeV. We will refer to this type of source as ``$\gamma$-ray-transparent".  

The diffuse extragalactic $\gamma$-ray background (EGB) has been measured by the {\it Fermi}-LAT between 100~MeV and 1~TeV. Depending on the energy, $\sim 30-80\%$ of the EGB is contributed by resolved sources and foreground emission \citep{FermiIGRB}. Above 50~GeV, $\sim 86\%$ of the EGB may be explained by unresolved blazars \citep{2016PhRvL.116o1105A}. The remaining part of the EGB is the isotropic $\gamma$-ray background (IGRB), which is composed of unresolved emissions, including the primary $\gamma$-rays from GeV-TeV protons (e.g., in low-mass, high-redshift starburst galaxies; \citealp{2021Natur.597..341R, 2022MNRAS.tmp.1040O}) and the cascades developed by TeV-PeV $\gamma$ rays and electrons of hadronic origin.

The origin of the diffuse $\gamma$-ray background in the MeV range is still largely unknown \citep{2016ApJ...820..142R}.  AGNs and Seyfert galaxies largely contribute to the diffuse emission from X-ray energies up to $\sim 0.3$~MeV, where their emission cuts off \citep{2003ApJ...598..886U}. Blazars, star-forming galaxies, and radio galaxies may account for the flux above $\sim 50-100$~MeV \citep{2015ApJ...800L..27A, 2015PhRvD..91l3001D}, but their contribution at a few MeV is expected to be $\lesssim 10\%$ \citep{2014ApJ...786...40L}. Different source models have been proposed for sources to fill the gap between 0.3 and 30~MeV, including the emission by non-thermal electrons in AGN coronae \citep{2013ApJ...776...33I}, MeV blazars \citep{2009ApJ...699..603A}, and radioactive nuclei in Type Ia supernovae \citep{2016ApJ...820..142R}. It has been noted that $\gamma$-ray cascades in AGN coronae may also contribute to the poorly constrained MeV background at the $\sim 10-30\%$ level  \citep{2019ApJ...880...40I, 2020PhRvL.125a1101M}.

Previous analyses  have suggested that if the sources of neutrinos are $\gamma$-ray-transparent, their TeV-PeV neutrino and $\gamma$-ray spectral index needs to be  $\gamma_{\rm astro}\lesssim 2.1-2.2$ in order not to exceed the observed IGRB \citep{2013PhRvD..88l1301M} and neutrino sources could be ``hidden" cosmic-ray accelerators \citep{2016PhRvL.116g1101M, 2020PhRvD.101j3012C}. In light of the latest measurements of the diffuse neutrino spectrum, we compute the spectra of cascades in the EBL for various injection models. We show that upcoming observations of the neutrino spectrum between $\sim 1-10$~TeV will decisively determine whether neutrino sources are mainly optically thick to high-energy $\gamma$-rays. We further show that in case the sources are indeed $\gamma$-ray-obscured, their electromagnetic cascades will appear at lower energies. The cascaded $\gamma$-rays may contribute to the MeV-GeV diffuse $\gamma$-ray background, unless the sources have an exceptionally strong magnetic field in which pairs mostly cool through synchrotron radiation or (and) a dense medium that absorbs the MeV-GeV $\gamma$-rays.   

A future observation of sub-TeV diffuse astrophysical neutrino can limit the sources of neutrinos to astrophysical objects with intense radiation fields. We review the electromagnetic cascades initiated by the $\gamma$-rays that accompany cosmic neutrinos in Section~\ref{sec:EMCascades}. We present the GeV-TeV diffuse $\gamma$-ray emission from $\gamma$-ray-transparent sources in Section~\ref{sec:transparent} and the MeV-GeV diffuse $\gamma$-ray emission from $\gamma$-ray-obscured sources in Section~\ref{sec:dark}. We conclude and discuss in Section~\ref{sec:discussion}.

\section{Electromagnetic Cascades}\label{sec:EMCascades}

Gamma-rays are inevitably emitted when high-energy neutrinos are produced. Hadronic interactions of cosmic rays produce charged and neutral pions, and possibly other mesons, which decay into neutrinos and $\gamma$-rays, respectively. Charged pions decay into neutrinos by the dominant process $\pi^\pm\rightarrow \mu^\pm\nu_\mu (\bar{\nu}_\mu)\rightarrow e^\pm\nu_e (\bar{\nu}_e)\bar{\nu}_\mu \nu_\mu$, and neutral pions decay into a pair of gamma rays, $\pi^0\rightarrow 2\gamma$. The fluxes of $\gamma$-rays and neutrinos produced by protons in the source environment are related by (e.g. \citealp{2013PhRvD..88l1301M})
\begin{equation}\label{eqn:E2dNdE_gamma_nu}
    E_\gamma^2 \frac{dN_{\rm inj}}{dE_\gamma} \approx \frac{4}{3 K_\pi } E_\nu^2 \frac{dN}{dE_\nu}  \Big| _{E_\nu = {E_\gamma}/2},
\end{equation}
where $K_\pi$ is the ratio of charged and neutral pions produced, with $K_\pi \approx 2 (1)$ for $pp (p\gamma$) interactions \citep{1996PhDT........59R, 2010ApJ...721..630H}.

In addition to the $\gamma$-rays originating directly from the decay of neutral pions, $\gamma$-rays may be produced leptonically, in particular, by the inverse Compton scattering by relativistic electrons. Gamma-rays from the decay of neutral pions thus represent the minimum energy in $\gamma$-rays from a neutrino-emitting source.

Gamma-rays interact with the source radiation field for sources that are $\gamma$-ray-obscured, or interact with the EBL upon leaving the source, in the case of $\gamma$-ray transparent sources. In both cases, the spectrum of the electromagnetic cascades has a universal shape that is independent of the spectral shape of the injected $\gamma$-rays, as first noticed by \citet{Berezinsky:1975zz, 1990acr..book.....B} and demonstrated in Appendix~\ref{sec:mono}. In addition, we find that the cascaded photon spectrum only weakly depends on the spectrum of the target radiation field; see Appendix~\ref{sec:AGN}. These two features make our study of the $\gamma$-ray cascades accompanying neutrinos independent of the modeling of the details of the source.

The radiation field in an astrophysical environment usually contains multiple components. For example, the emission from the inner region of AGN can be described by a two-phase model, which includes thermal UV emission by the disk and hard non-thermal X-ray emission by the corona. Let us assume that the lowest- and the highest-energy $\gamma$-ray attenuating component of the radiation field have energies  $\varepsilon_l$ and $\varepsilon_h$, respectively. We further assume that the photon number density of the low-energy component is higher than that of the high-energy one, as is the case for most astrophysical sources.  

The energy spectrum of the cascades peaks at the cutoff energy that corresponds to the pair production threshold with the highest-energy background photons, 
\begin{equation}\label{eqn:Eg_di}
    {\cal E}_\gamma = \frac{4 m_e^2}{\varepsilon_h}. 
\end{equation}
The prefactor $4$ comes from the fact that the pair production cross section peaks at the center-of-mass energy squared $X=2m_e^2$, where $X = E_\gamma \varepsilon (1- \mu)/2\sim E_\gamma \varepsilon/2$ is the Lorentz invariant interaction energy, assuming an average interaction angle $\mu=0$.

The inverse-Compton emission by the last generation of electrons from the pair production yields a second peak at ${\cal E}_X$,  which is dominated by the low-energy radiation field, 
\begin{equation}\label{eqn:Ex_di}
    {\cal E}_X = \frac{4}{3} \left(\frac{{\cal E}_\gamma}{2 m_e}\right)^2 \varepsilon_l.
\end{equation}
The cascade spectrum follows a power law $dN_{\rm cas}/dE_\gamma\propto E_\gamma^{-3/2}$ below ${\cal E}_X$, $dN_{\rm cas}/dE_\gamma\propto E_\gamma^{-1.9}$ between ${\cal E}_X$ and ${\cal E}_\gamma$, and cuts off above ${\cal E}_\gamma$. 

As further explained in Appendix~\ref{sec:AGN}, although a source radiation field may have a broad spectral energy distribution, ${\cal E}_\gamma$ and ${\cal E}_X$ are determined by the highest- and lowest-energy background photons that are optically thick to $\gamma$-rays.

Finally, since the total energy of the cascades is conserved \citep{1990acr..book.....B}, the flux of the cascades is determined by the injected $\gamma$-ray power as,  
\begin{equation}
    \int dE_{\gamma} E_{\gamma}\frac{dN_{\rm cas}}{dE_{\gamma}} = \int_{E_{\gamma, \rm min}^{\rm inj}} dE_{\gamma}  E_{\gamma} \frac{dN_{\rm inj}}{dE_{\gamma}}, 
\end{equation}
which, notably, does not depend on the shape of $dN_{\rm inj}/dE_\gamma$ as long as the injected $\gamma$ rays are fully attenuated, that is, $E_{\gamma\rm min}^{\rm inj} > {\cal E}_\gamma$.

\section{Gamma-ray-Transparent Neutrino Sources}\label{sec:transparent}

\begin{figure*}[t]
    \centering
   \includegraphics[width=0.49\textwidth]{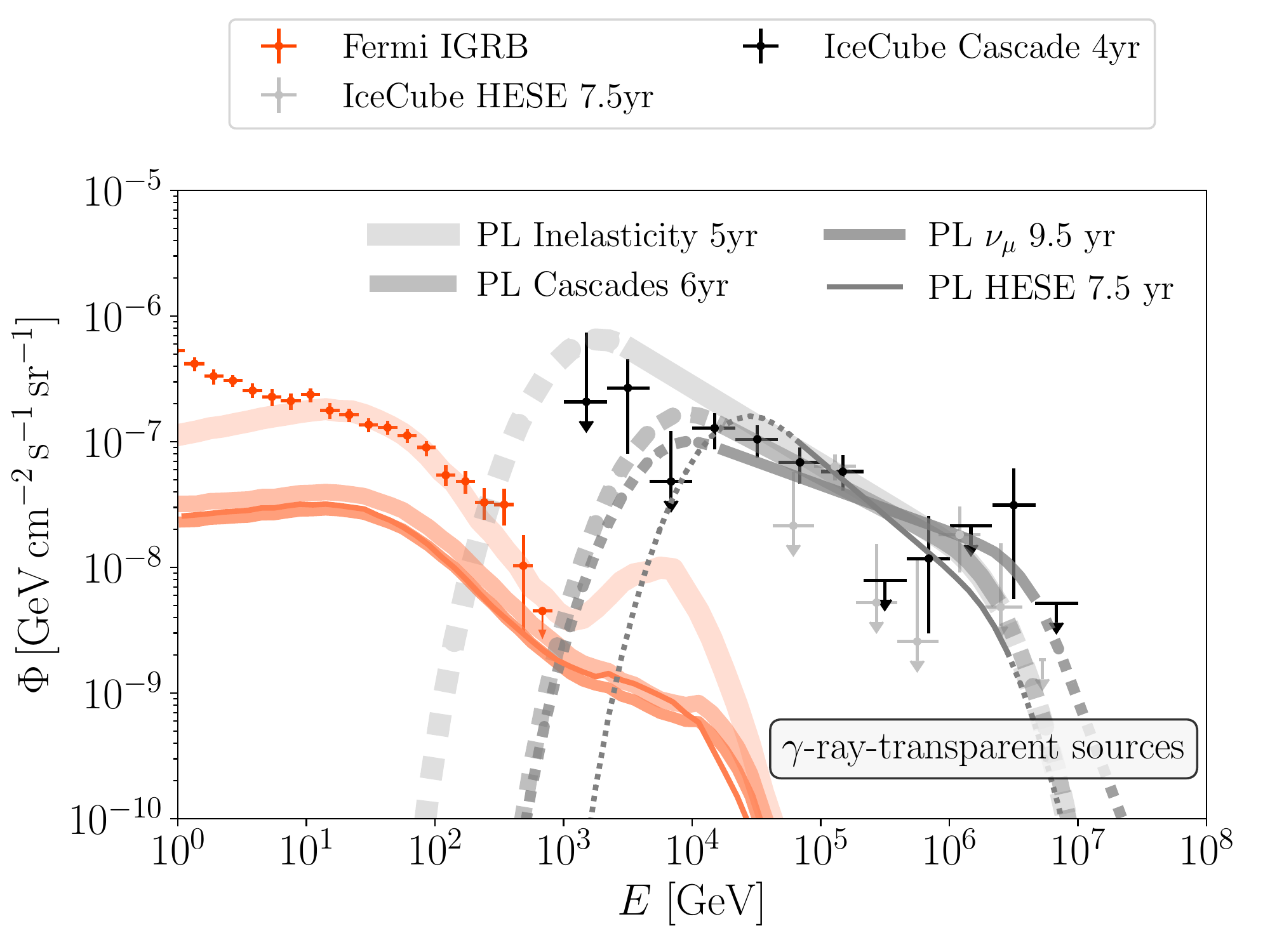}
      \includegraphics[width=0.49\textwidth]{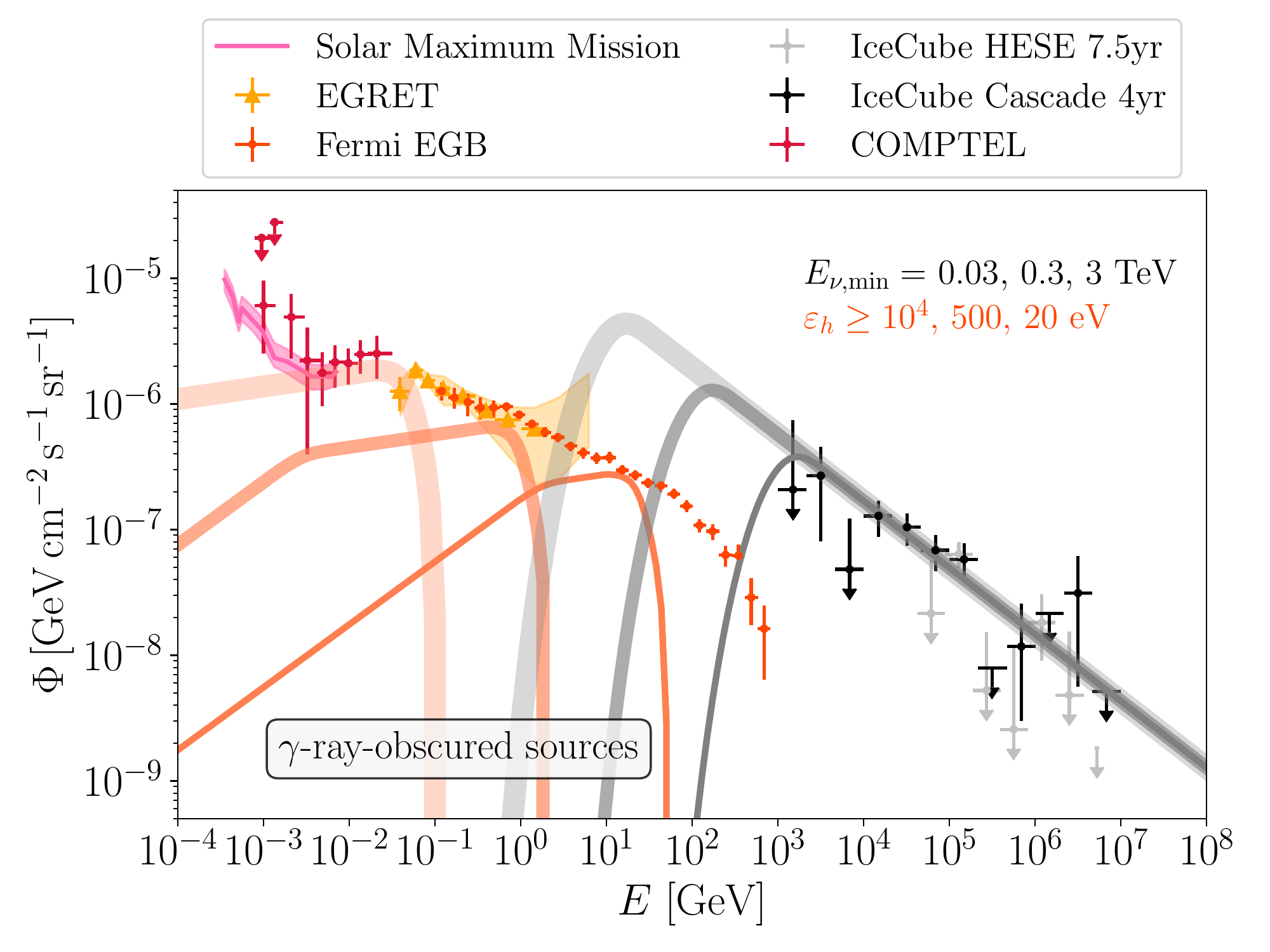}
    \caption{\label{fig:sed} Cosmic neutrino spectra (black curves) and $\gamma$-ray cascades initiated by their electromagnetic counterparts (red curves). The data points are measurements of the diffuse cosmic neutrino background \citep{2017arXiv171001191I, HESE75yr}, extragalactic $\gamma$-ray background (EGB) and isotropic $\gamma$-ray background (IGRB) (\citealp{FermiIGRB} assuming foreground model A)  from 0.1~GeV to 1~TeV, and diffuse MeV $\gamma$-ray background  \citep{1999PhDT.......284W,1999ApJ...516..285W, 2004ApJ...613..956S}. {\bf Left: $\gamma$-ray-transparent sources.} Neutrino spectra correspond to the best-fit single power-law models from the observations listed in Table~\ref{table:parameters}. Fluxes below and above the sensitivity range for the IceCube analyses are unknown and shown as dashed curves.  Gamma-rays from hadronic interactions leave the sources without attenuation, propagate in the extragalactic background light (EBL), and cascade down to GeV-TeV energies. Note that the {\it Fermi} IGRB may also be contributed by additional emission mechanisms such as inverse Compton scattering by relativistic electrons accelerated by sources, and thus serves as an upper limit on the cascaded hadronic $\gamma$-rays.  {\bf Right: Gamma-ray-obscured sources.} Neutrino spectra are assumed to follow single power-law models with the best-fit index from the cascade neutrino sample \citep{cascade6yr} and a minimum cutoff energy at 0.03, 0.3, and 3~TeV, respectively. A source radiation field at or above hard X-ray, soft X-ray, and UV energies, correspondingly, is needed to attenuate the hadronic $\gamma$ rays from each cutoff energy such that the cascades flux is below the MeV-GeV diffuse $\gamma$-ray background. In either scenario for $\gamma$-ray transparency, the sub-TeV neutrino spectrum encodes crucial information about the cosmic environment of the astrophysical sources of high-energy neutrinos.  
    } 
\end{figure*}

We first consider the scenario where the neutrino sources are transparent to $\gamma$-rays. In this case, the $\gamma$-rays produced in association with the neutrinos leave the source without interacting and losing energy; they subsequently propagate in the EBL to our detectors. The injection spectrum of $\gamma$-rays into the EBL is obtained from the latest measurements of the diffuse neutrino spectrum observed by IceCube using equation~\ref{eqn:E2dNdE_gamma_nu}. The neutrino spectrum is parameterised as 
\begin{equation}\label{eqn:phinu}
  \frac{dN}{dE_\nu}  \propto E_\nu^{-\gamma_{\rm astro}},\, E_{\nu, \rm min} \leq E_\nu \leq E_{\nu, \rm max}.
\end{equation}
The spectral index $\gamma_{\rm astro}$ of the neutrino spectrum is set to the best-fit parameters found by \citet{HESE75yr, cascade6yr, numu95yr, inelasticity5yr} when fitting the astrophysical neutrino flux with a single power-law. The minimum and maximum energies are set to the range of the neutrino energies that the particular analysis is sensitive to. The values of the parameters are summarized in Table~\ref{table:parameters}.   

Outside the measured range, we invoke a conservative exponential cutoff of the neutrino spectrum below $E_{\nu, \rm min}$ and above $E_{\nu, \rm max}$. Specifically, 
\begin{equation}\label{eqn:cutoff}
    \frac{dN}{dE_\nu}=\begin{cases}
			\frac{dN}{dE_\nu}(E_{\nu,\rm min}) e^{ 1 - E_{\nu,\rm min}/E_\nu }, &    E_\nu < E_{\nu,\rm min} \\
            \frac{dN}{dE_\nu}(E_{\nu,\rm max}) e^{1 - E_\nu / E_{\nu,\rm max}}, &    E_\nu > E_{\nu,\rm max}
		 \end{cases}
\end{equation}
The actual, thus far unobserved neutrino spectrum, could extend well beyond these cutoff energies. 

\begin{table*}
\caption{Summary of single power-law parameters fitted to the measurements of the cosmic neutrino flux used in equation~\ref{eqn:phinu}. All measurements assume an equal flux of neutrinos and antineutrinos, and an equal flux of the three neutrino flavors.    \label{table:parameters}}
\begin{ruledtabular}
\begin{tabular}{cccccc}
  Dataset  &  $\Phi_{\rm astro}$  & $\gamma_{\rm astro}$  & $E_{\nu, \rm min}$  &  $E_{\nu, \rm max}$ & reference     \\ 
  &  [$3\times 10^{-18}\,\rm GeV^{-1}cm^{-2}s^{-1}sr^{-1}$]   & &  [TeV]  &  [PeV]  &    \\
  \hline
 HESE 7.5 years   \space  & 2.12   & 2.87  & 60 & 3     &  \citet{HESE75yr}    \\
 Cascades 6 years  \space     & 1.66   & 2.53 & 16 & 2.6 &  \citet{cascade6yr} \\
 $\nu_\mu$ 9.5 years  \space     & 1.44  & 2.37 & 15 & 5 &  \citet{numu95yr} \\
 Inelasticity 5 years  \space     & 2.04   & 2.62 & 3.5 & 2.6 &  \citet{inelasticity5yr} \\
\end{tabular}
\end{ruledtabular}
\end{table*}

We assume that the neutrinos are produced via $pp$ interactions, because such sources will be $\gamma$-ray-obscured in a $p\gamma$ scenario given that the cross section of $p\gamma$ interactions is smaller than than that for $\gamma\gamma$ interactions \citep{{2016PhRvL.116g1101M}}. 

Contributions of neutrino sources from different redshifts are integrated, taking into account the energy loss due to cosmological expansion (e.g., \citealp{2006ApJ...651..142H}):
\begin{equation}
    \Phi_\nu (E_\nu) = \frac{c}{4\pi} \int_{z_{\rm min}}^{z_{\rm max}} dz  \Big|\frac{dt}{dz}\Big|  (1+z) \dot{\rho}_{\rm sr}(z) \frac{dN (E_\nu'=E_\nu(1+z))}{dE'_\nu}
\end{equation}
where $|dt/dz| = (H_0 (1+z) \sqrt{\Omega_M (1+z)^3 + \Omega_\Lambda})^{-1}$, $\dot{\rho}_{\rm sr} (z)= \dot{\rho}_{\rm sr}(z=0) g(z)$ is the source emissivity, $\dot{\rho}_0$ is the local source emissivity in units of $\rm Mpc^{-3}\,yr^{-1}$, and $g(z)$ denotes the relative source evolution rate over redshift. We adopt a standard $\Lambda$CDM flat cosmology with $\Omega_M=0.315$ \citep{2020A&A...641A...6P}. Our calculation assumes that the source distribution follows the star-formation (SFR) history of the universe \citep{2006ApJ...651..142H}, though the impact of the source history model on the integrated cascade spectrum is relatively small \footnote{For reference, the diffuse $\gamma$-ray flux above $\sim 300$~GeV in Figure~\ref{fig:sed}-left, which mainly comes from nearby sources, may increase by a factor of $\sim 2-3$ if the source emissivity is uniform, $g(z)=1$. The flux below $\sim 300$~GeV is similar. In an extreme scenario where most sources are at $z>4$ (e.g. discussed in \citealp{Xiao:2016rvd}), the diffuse $\gamma$-ray flux above $\sim 300$~GeV would be $\sim 0$. }. The electromagnetic cascades are computed numerically using the EBL model in \citet{2011MNRAS.410.2556D} integrated between $z_{\rm min}=0.001$ and $z_{\rm max}=4$. The differential neutrino flux is normalized to observations at 100~TeV, $\Phi_\nu(E_\nu = 100 \,\rm TeV) = \Phi_{\rm astro}$, with $\Phi_{\rm astro}$ summarized in Table~\ref{table:parameters}.

The left panel of Figure~\ref{fig:sed} presents the diffuse neutrino flux and the electromagnetic cascades from their $\gamma$-ray counterparts. As the neutrino flux magnitudes and spectral indices from the four measurements are comparable, the flux of the cascades is determined by $E_{\nu,\rm min}$. When $E_{\nu,\rm min}\gtrsim 10$~TeV, the cascades contribute up to $\sim 30-50\%$ of the IGRB between 30~GeV and 300~GeV, and nearly 100\% above 500~GeV. The cascade flux would exceed the IGRB above $\sim 10$~GeV when $E_{\nu,\rm min}\lesssim 5$~TeV assuming that the measured power law distribution continues. 

The constraints from the IGRB may be tighter than what is shown in Figure~\ref{fig:sed} for two reasons. First, the cascade flux only includes $\gamma$ rays from the hadronic processes. If the magnetic field of the neutrino-emitting region is not strong, electrons from the decay of charged pions may also produce $\gamma$ rays that contribute to the cascades in the EBL.  Gamma-rays from relativistic electrons accelerated in the sources, which could be comparable or even dominate over those from protons, would further increase the cascade flux. Second, the IGRB attributed to the neutrino counterparts may be much lower than the $\it Fermi$-LAT data points in use, since a large fraction of the IGRB is known to be contributed by the primary GeV-TeV emission by the unresolved, low-photon-count extension of known sources \citep{2016PhRvL.116o1105A, 2021Natur.597..341R, 2022MNRAS.tmp.1040O}. In the end, the room left to accommodate secondary photons from TeV-PeV $\gamma$ rays and electrons is small.

\section{$\gamma$-ray-obscured Neutrino Sources}\label{sec:dark}

In the emerging scenario where most neutrinos sources are optically thick to high-energy $\gamma$-rays, electromagnetic cascades develop inside the sources. The energy carried by the neutrino counterparts, therefore, will show up at lower energies. The peak energy of the cascade spectrum is determined by the characteristic photon energy of the radiation field that interacts with the $\gamma$-rays (see Appendix~\ref{sec:AGN}). Because the flux of the cascades may not exceed the measured diffuse $\gamma$-ray background, constraints on the energy and density of the source radiation field can be obtained. 

As a demonstration, the right panel of Figure~\ref{fig:sed} shows the diffuse neutrino and cascade spectra from $\gamma$-ray-obscured sources assuming $E_{\nu,\rm min} = 0.03$, 0.3, and 3~TeV, respectively. In all three cases, the neutrino spectrum is assumed to be a single power-law with $\gamma_{\rm astro}=2.53$, motivated by the index measured with the IceCube cascade cosmic neutrino sample \citep{cascade6yr}. As $E_{\nu,\rm min}$ decreases, the neutrinos and injected $\gamma$ rays carry more power, and a source radiation field with higher $\varepsilon_h$ is needed to reprocess the $\gamma$-ray power to lower energies to be consistent with the $\gamma$-ray observations. 

Without loss of generality, we assume that the number density of the source radiation field, $\varepsilon dn/d\varepsilon$, peaks at $\varepsilon_l = 1$~eV. We again assume that the source emissivity follows the star-forming history of the Universe.  A minimal $\varepsilon_h$ may then be obtained such that the integrated flux of the cascades is below the EGB \footnote{Although a neutrino-emitting site can be $\gamma$-ray-obscured, the source may produce $\gamma$ rays via leptonic processes from a different region. As the $\gamma$-ray production sites may not be resolved by $\gamma$-ray telescopes, the EGB poses a more conservative upper limit than the IGRB on the integrated cascade flux.}. The right panel of Figure~\ref{fig:sed} shows that the currently measured neutrino spectrum with $E_{\nu, \rm min} \sim 3$~TeV already requires a source radiation field populated with UV photons at $\varepsilon_h \gtrsim 20$~eV. If future observations find an even lower $E_{\nu, \rm min}$, an intense X-ray radiation field must be present in the neutrino production site to produce the required optical depth. 

The value of $\varepsilon_h$ only weakly depends on $\varepsilon_l$ since the energy flux of cascades peaks at ${\cal E}_\gamma \propto \varepsilon_h^{-1}$. As explained in Appendix~\ref{sec:AGN}, the cascade spectrum is mostly universal with respect to the spectrum of the radiation field. {\it The form of $\varepsilon_h$ and the conclusion that the sources must be $\gamma$-ray-obscured at such an energy are essentially  model-independent. }

\section{Conclusions and Discussion}\label{sec:discussion}

High-energy neutrinos are inevitably accompanied by a flux of $\gamma$-rays. Unlike neutrinos which barely interact, $\gamma$-rays pair produce with radiation fields, either inside the neutrino source or propagating through the EBL, reprocessing their power to lower energies. Because the flux and spectral index of the TeV-PeV diffuse neutrino background are comparable to that of the GeV-TeV diffuse $\gamma$-ray background, the latter tightly constrains the flux of the electromagnetic cascades of the $\gamma$-ray counterparts of high-energy neutrinos. By comparing the flux of these cascades developed in the EBL to the IGRB, we show that the current IceCube measurements already indicate that the bulk of the neutrino sources are likely opaque to $\gamma$-rays. Assuming that the sources are $\gamma$-ray-obscured, we find that the MeV-TeV diffuse $\gamma$-ray background confines the minimum energy of the source's radiation field in which pair production must effectively happen, and thus the total power in the sources. 

Future improved measurement of the neutrino spectrum at lower energies will 1) unambiguously confirm that the sources of high energy neutrinos are optically thick to GeV-TeV $\gamma$-rays, and 2) suggest that the neutrino sources contain intense high-energy (soft X-ray or higher) internal radiation backgrounds. These conclusions are independent of the modeling of the source because there is a direct link between the production rates of neutrinos and $\gamma$-rays and because of the universality of the spectrum of the electromagnetic cascades.

The cascade flux calculation in the left panel of Figure~\ref{fig:sed} does not account for the effect of intergalactic magnetic field (IGMF). The effect of IGMF on the cascade spectrum above 10~GeV, where most constraints come from, is expected to be minor for $B_{\rm IGMF} \lesssim 10^{-13}$~G \citep{2013MNRAS.432.3485V}. Such a field strength is consistent with constraints found by stacking of pair halos around distant AGNs \citep{2021Univ....7..223A}.

We did not consider the effect of magnetic fields on the development of cascades in neutrino sources. This is because the energy density of the photon fields is expected to exceed the magnetic energy density in many promising candidate sources such as the AGN coronae \citep{2019ApJ...880...40I, 2020PhRvL.125a1101M}, tidal disruption events \citep{2021NatAs...5..510S}, and shock-powered optical transients \citep{2020ApJ...904....4F}. The presence of an exceptionally intense magnetic field in combination with a low radiation field in the source could impact the development of the electromagnetic cascades. The flux of cascades would be lower since the energy of the pairs is dissipated through synchrotron radiation. We also did not consider the absorption of MeV-GeV $\gamma$-rays, which could happen if a source has a high-density matter field, where high-energy photons interact with free electrons through Compton scattering or (and) protons and nuclei through the Bethe-Heitler process. Such absorption could happen for example when a particle accelerator is embedded in a stellar ejecta (e.g., \citealp{2013PhRvL.111l1102M, 2020ApJ...904....4F}). As long as the cascades developed in the neutrino-emitting site may leave the source, the cascaded flux is uniquely linked to the neutrino flux even when a source has more than one emission zone. 


Neutrino production may occur in a relativistic flow. Assuming that the plasma has a bulk Lorentz factor $\Gamma$, the source target photon field needs to have $\varepsilon_h = 4 \Gamma^2 m_e^2/ {\cal E}_\gamma$, since the injected $\gamma$-ray energy is $\Gamma$ times lower than the observed value in the rest frame of the plasma. The need of an intense high-energy radiation field in the to block $\gamma$-rays from neutrino sources is thus even more severe if neutrinos are from regions moving with relativistic speed.  

Near-term observations by IceCube \citep{2021arXiv210709811M} may measure the neutrino spectrum at 1--10~TeV. Future observations by IceCube-Gen2 \citep{2021JPhG...48f0501A}, KM3NeT \citep{2016JPhG...43h4001A}, and Baikal-GVD \citep{2021arXiv210801894A} may extend to the sub-TeV regime and unveil the nature of the neutrino sources. These observations will place fundamental limits on the nature of cosmic sources of high energy neutrinos.

\vspace{2em}

\begin{acknowledgments}
We thank Markus Ahlers, Albrecht Karle, and Kohta Murase for helpful comments on the
manuscript. The work of K.F and F.H is supported by the Office of the Vice Chancellor for Research and Graduate Education at the University of Wisconsin-Madison with funding from the Wisconsin Alumni Research Foundation. K.F. acknowledges support from NASA (NNH19ZDA001N-Fermi, NNH20ZDA001N-Fermi, NNH20ZDA001N-Swift) and National Science Foundation (PHY-2110821). J.S.G. thanks the University of Wisconsin College of Letters and Science for partial support of his IceCube-related research. The research of F.H was also supported in part by the U.S. National Science Foundation under grants~PLR-1600823 and PHY-1913607.

\end{acknowledgments}

\appendix
 
\section{Analytical Estimation of Cascade Spectrum}\label{sec:mono}

Electromagnetic cascades have been well studied in the context of high-energy $\gamma$-ray propagation in the extragalactic background light (EBL). The cascade spectrum can be described as \citep{1990acr..book.....B, 2016PhRvD..94b3007B}: 
\begin{equation}\label{eqn:analySpectrum}
\frac{dN_{\rm cas}}{dE_\gamma}=\begin{cases}
			K  (E_\gamma/{\cal E}_X)^{-3/2}, &    E_\gamma < {\cal E}_X \\
             K  (E_\gamma/{\cal E}_X)^{-2}, &  {\cal E}_X < E_\gamma < {\cal E}_\gamma \\
            0, & E_\gamma > {\cal E}_\gamma
		 \end{cases}
\end{equation}
where ${\cal E}_\gamma$ corresponds to the threshold $\gamma$-ray energy for pair production, and ${\cal E}_X$ is the $\gamma$-ray energy of the photons up-scattered by the last-generation of electron positron pairs. The transition from the $E_\gamma^{-2}$ to $E_\gamma^{-3/2}$ regimes happens when no new pairs are produced and the number of pairs stays constant.

The energy carried by the primary $\gamma$-rays and pairs, 
\begin{equation}
    W_{\rm inj} \equiv \int dE_\gamma E_\gamma \frac{dN_{\rm inj}}{dE_\gamma},
\end{equation}
is transferred to the cascades. The prefactor can be derived from energy conservation, 
\begin{equation}\label{eqn:analyNorm}
    K = \frac{W_{\rm inj}}{{\cal E}_X^2}\left(2+\ln\frac{{\cal E}_\gamma}{{\cal E}_X}\right)^{-1}. 
\end{equation}

The energy flux of the cascades peaks at ${\cal E}_\gamma$, $(E^2_\gamma dN/dE_\gamma)^{\rm pk}_{\rm cas} \approx K {\cal E}_X^2$. In case that the injection spectrum follows a steep power-law, $(dN/dE_\gamma)_{\rm inj}\propto E_\gamma^{-s}$ with $s> 2$, the peak energy fluxes of the cascades and injected photons are related by 
\begin{equation}
    \left(E^2_\gamma \frac{dN_{\rm cas}}{dE_\gamma}\right)_{\rm pk} \approx \eta \, \left(E^2_\gamma \frac{dN_{\rm inj}}{dE_\gamma}\right)_{\rm pk},
\end{equation}
where 
\begin{equation}
    \eta \equiv \frac{1}{s-2}  \left(2+\ln\frac{{\cal E}_\gamma}{{\cal E}_X}\right)^{-1}
\end{equation}
is a factor of order unity. This is why in an energy flux ($E^2dN/dE$) plot, cascades appear to inherit the flux of the injected $\gamma$ rays at a lower energies.

\begin{figure} 
    \centering
   \includegraphics[width=0.49\textwidth]{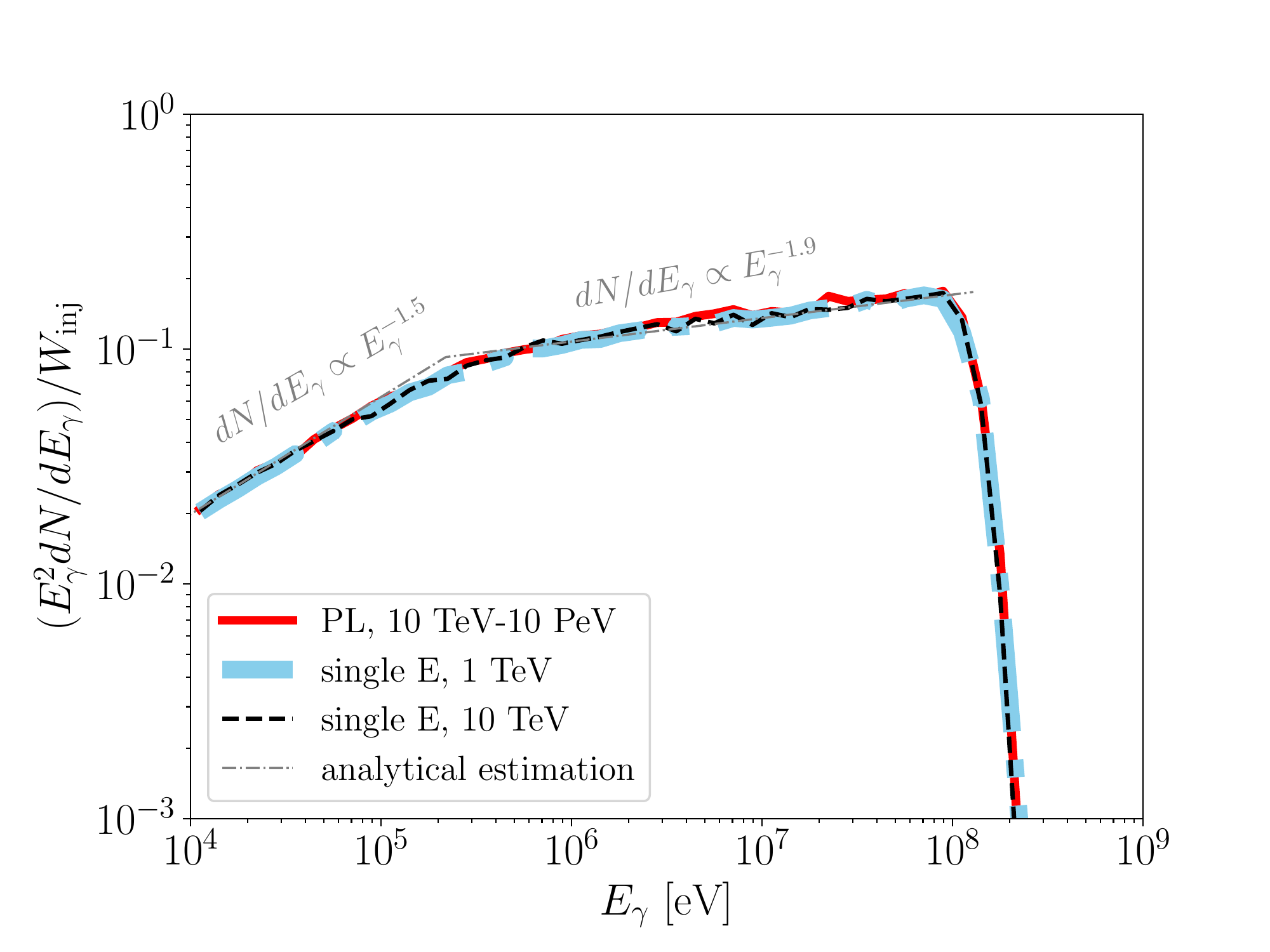}
    \caption{\label{fig:dichromatic} {\bf Illustration of the universality of the cascade spectrum with respect to the spectrum of the injected $\gamma$-rays.} Energy spectra of electromagnetic cascades of $\gamma$-rays from various injection models in a dichromatic radiation background are  normalized by the injected energy $W_{\rm inj}$. Three injection models are in use: 1) $\gamma$-rays follow a power-law spectrum between 10~TeV and 10~PeV with a spectral index $s = 2.5$ as motivated by the IceCube Cascade sample, 2) all $\gamma$-rays have the same energy of 1~TeV, and 3) all $\gamma$-rays have the same energy of 10~TeV. Cascade spectrum does not depend on the shape of the $\gamma$-ray injection spectrum.  
    } 
\end{figure}

\begin{figure} 
    \centering
   \includegraphics[width=0.49\textwidth]{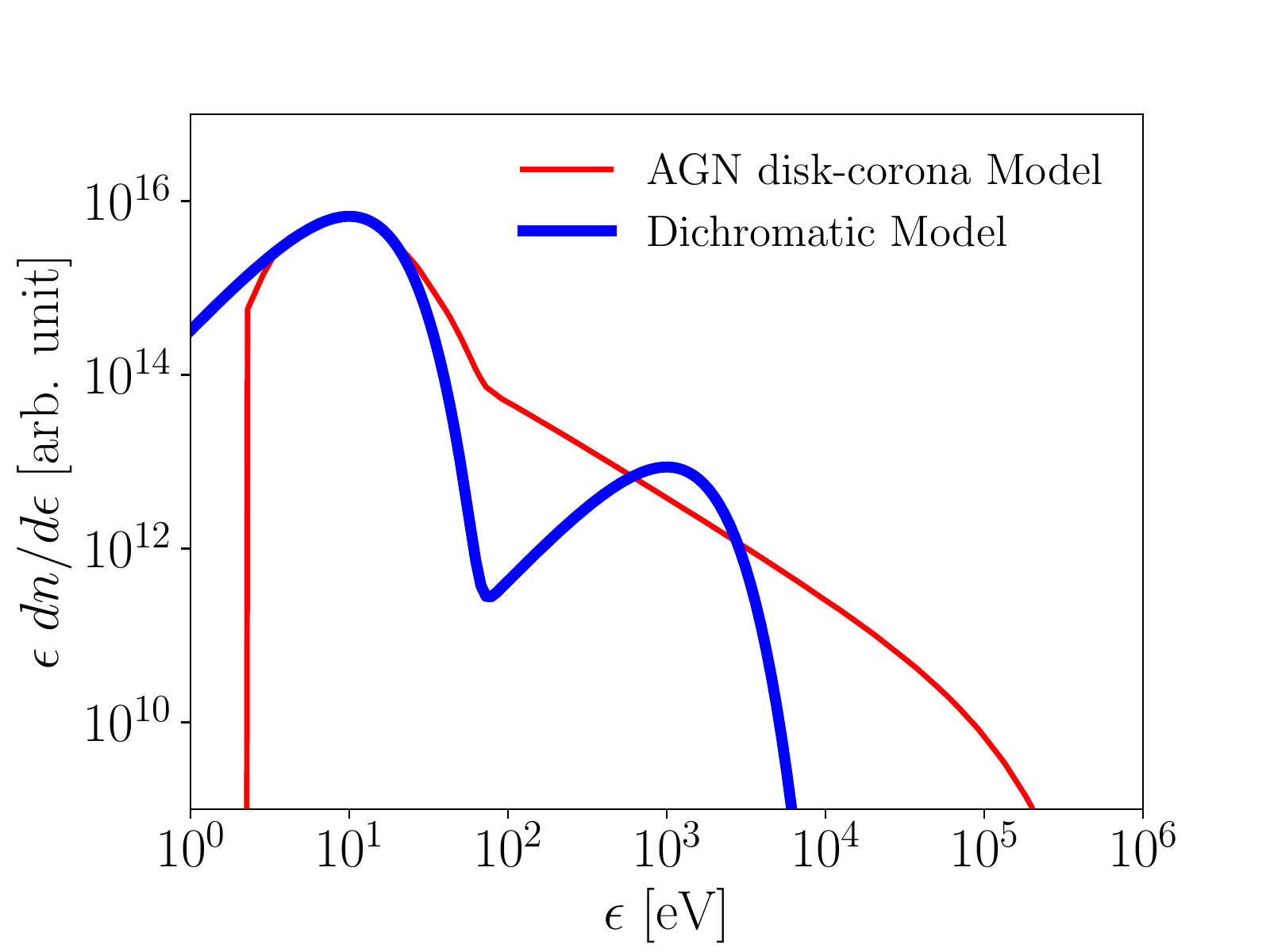}
   \includegraphics[width=0.49\textwidth]{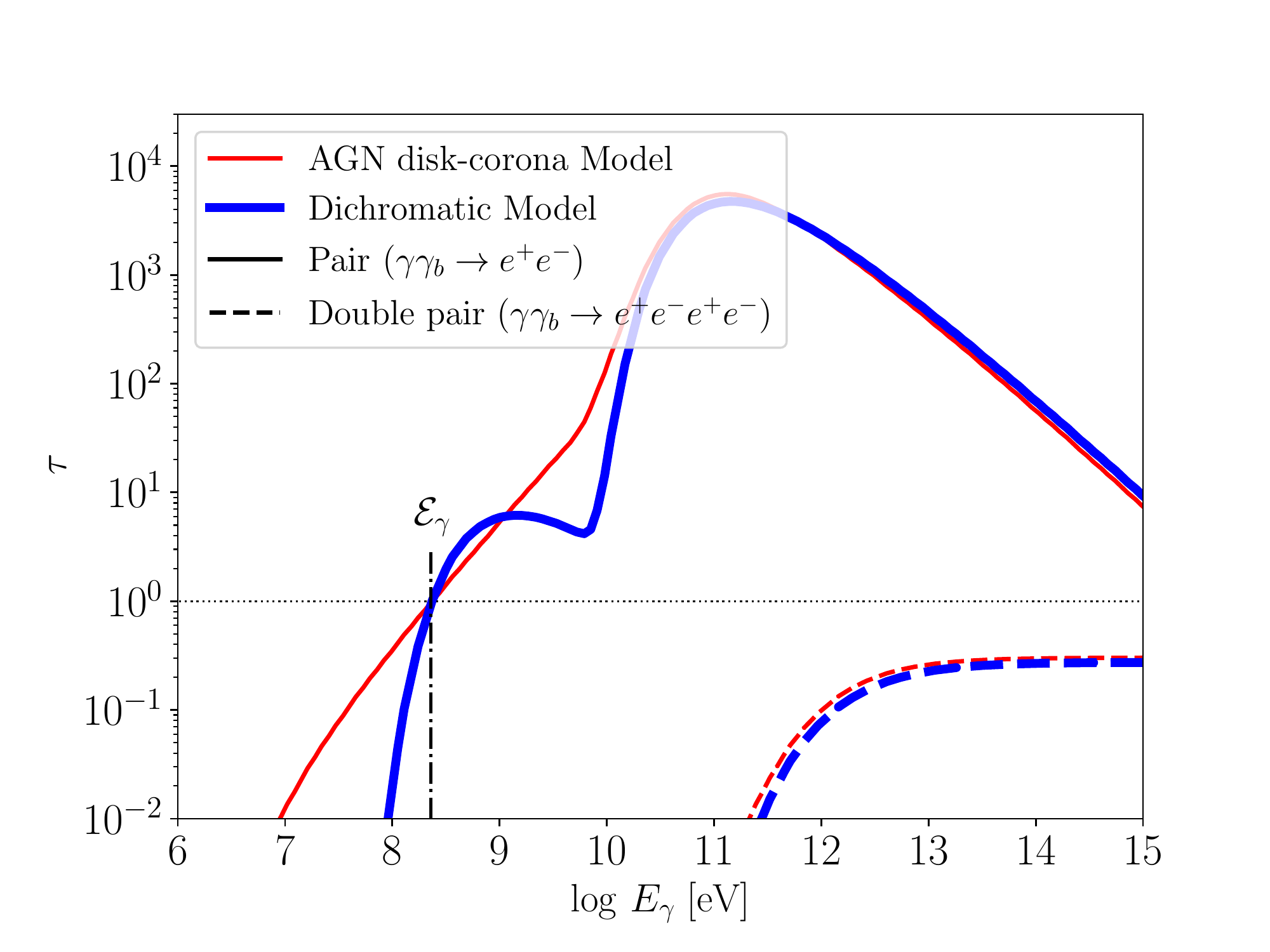}
   \includegraphics[width=0.49\textwidth]{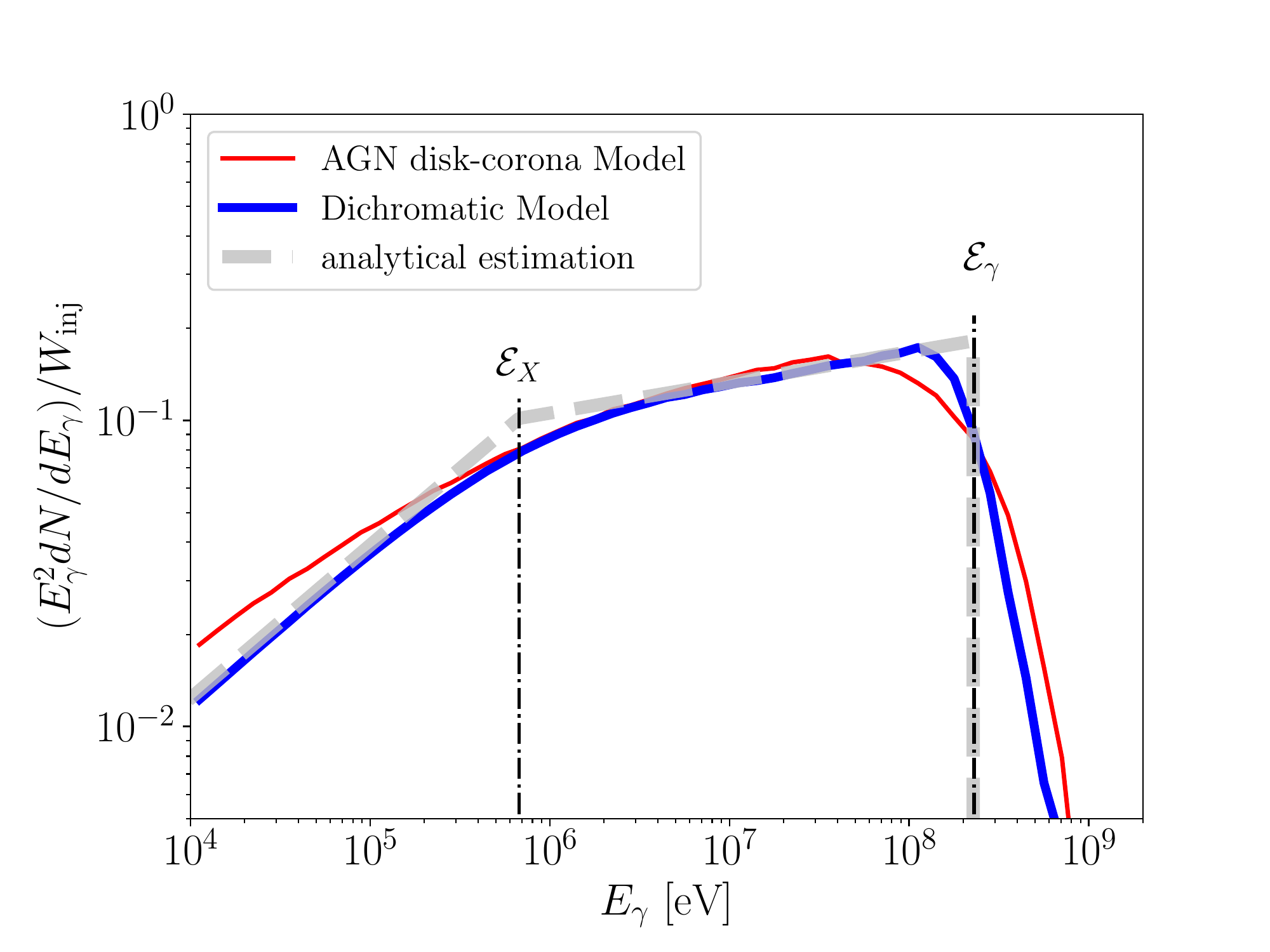}
    \caption{\label{fig:AGN} {\bf Illustration of the universality of the cascade spectrum with respect to the spectrum of the target radiation field.} {\bf Top:} Spectra of an example AGN disk-corona emission model and a dichromatic radiation field. {\bf Middle:} Optical depth $\tau$ due to pair (solid curves) and double pair (dashed curves) production. Optical depth is set to $\tau=1$ at $E_\gamma={\cal E}_\gamma$ for both fields. {\bf Bottom:} The corresponding cascade spectra. The grey dashed curve shows the analytical spectrum calculated by equation~\ref{eqn:analySpectrum}  with the ${\cal E}_\gamma$ in the middle panel and ${\cal E}_X$ defined by equation \ref{eqn:Ex_di}. In all panels, a red thin curve indicates the AGN model and a blue thick curve indicates the dichromatic model. 
    } 
\end{figure}

The spectral shape of the cascades results from the distribution of energy over particle generations and is universal for given ${\cal E}_\gamma$,  ${\cal E}_X$, and $W_{\rm inj}$. It does not depend on the spectral shape of the injected $\gamma$ rays and electrons.

Figure~\ref{fig:dichromatic} presents the cascade spectra of various injection models in a dichromatic field with peak energies $\varepsilon_{l}^{\rm di} = 10$~eV and $\varepsilon_{h}^{\rm di} = 1$~keV. 
The cascade spectra are normalized by the injected energy $W_{\rm inj}$. 

The calculation is performed numerically with the CRPropa package \citep{2016JCAP...05..038A} with customized photon fields and interaction tables \footnote{A problem with the interaction tables of the CRPropa code (version 3.1.7) was noticed by \citet{2022arXiv220103996K}. We have corrected the interaction rate calculation accordingly.}. The interaction modules include pair and double pair production of $\gamma$-rays, and inverse Compton scatter and triplet pair production of relativistic electrons. The development of secondary particles is tracked down to $E_\gamma = 10^4$~eV. 

As shown in Figure~\ref{fig:dichromatic}, the spectra of cascade $\gamma$-rays from different injection models are similar. The fluxes and break energies are fully determined by the injection energy and the radiation field, respectively. They are not sensitive to the injected $\gamma$-ray spectrum. We confirm the findings of \citet{2016PhRvD..94b3007B} that the intermediate component of the cascade spectrum is better described by $E^{-1.9}$ than $E^{-2}$.

\section{Electromagnetic Cascade in AGN Cores}\label{sec:AGN}

In this section, we investigate the dependence of the cascade spectrum on the spectral energy distribution of the target radiation field. Electromagnetic cascades are calculated with two different types of radiation fields: 1) the benchmark dichromatic field used in Appendix~\ref{sec:mono}, and 2) a typical AGN disk-corona emission model which includes a thermal emission component from the accretion disk that peaks at $\varepsilon_{l}^{\rm AGN} =10$~eV, a soft X-ray excess, and characteristic coronal power-law emission \citep{2017MNRAS.465..358C}. The number density $dn/d\varepsilon$ of the two radiation fields is shown in the upper panel of Figure~\ref{fig:AGN}.  

For an isotropic photon field, the absorption probability per unit path length is \citep{2009herb.book.....D}
\begin{equation}
    \frac{d\tau_{\gamma\gamma}}{dx} (E_\gamma) = \frac{1}{2}\int_{-1}^{1}d\mu (1-\mu) \int_0^\infty d\varepsilon \frac{dn}{d\varepsilon}(x)\sigma_{\gamma\gamma}(y),
\end{equation}
where $y = E_\gamma \varepsilon (1-\mu) / (2 m_e^2)\equiv \gamma_{\rm cm}^2$, $\gamma_{\rm cm}$ is the Lorentz factor of the produced pairs in the center-of-momentum frame, and $\sigma_{\gamma\gamma}$ is the cross section of pair production, 
\begin{eqnarray}
    \sigma_{\gamma\gamma} (y) &=& \frac{3}{16}\sigma_T (1 - \beta_{\rm cm}^2) \\ \nonumber
    &\times& \left[(3- \beta_{\rm cm}^4) \ln \left(\frac{1 + \beta_{\rm cm}}{1-\beta_{\rm cm}}\right) - 2 \beta_{\rm cm}(2-\beta_{\rm cm}^2)\right]
\end{eqnarray}
with $\beta_{\rm cm} = (1 -\gamma_{\rm cm}^{-2})^{1/2}$. 

The optical depth for $\gamma$-rays at energy $E_\gamma$ of an interaction region $R$ can be written as 
\begin{equation}
    \tau_{\gamma\gamma} = \int_0^R dx \frac{d\tau_{\gamma\gamma}}{dx}.
\end{equation}

The middle panel of Figure~\ref{fig:AGN} presents the optical depth $\tau$ of the pair and double pair production interactions for both radiation fields. The number of expected interactions clearly tracks the spectral shape of the target field, as $\tau(E_\gamma)\approx n(\varepsilon = 4m_e^2 / E_\gamma)\sigma_{\gamma\gamma}R$. 

We set the optical depth of both fields to $\tau({\cal E}_\gamma)=1$ at a random energy ${\cal E}_\gamma$. The corresponding highest-energy target photons that may attenuate these $\gamma$ rays are at $\varepsilon_{h} \equiv 4 m_e^2 / {\cal E}_\gamma$. Note that $\varepsilon_{h}$ can be different from the high-energy peak of the dichromatic field $\varepsilon_{h}^{\rm di}$ and is determined by the optical depth. Because $\varepsilon dn/d\varepsilon$ peaks at $\varepsilon_l^{\rm AGN} = \varepsilon_l^{\rm di}$, ${\cal E}_X = ({\cal E}_\gamma / 2 m_e)^2 \varepsilon_{l}$ is the same for both fields. 

The spectra of cascades developed in the two fields are shown in the bottom panel of Figure~\ref{fig:AGN}. Despite the fact that the radiation fields have different spectral energy distributions, the resulting cascades present very similar spectra. They may be reasonably described by the analytical formula in equation~\ref{eqn:analySpectrum} normalized by equation~\ref{eqn:analyNorm}. The slight departure from $E_\gamma^{-3/2}$ below ${\cal E}_X$ in the AGN case is due to the spread in the seed photon energy of the inverse Compton scattering.

\bibliography{references} 
\bibliographystyle{aasjournal}

\end{document}